\documentclass[a4paper, 12pt]{article}

\makeatletter
\renewcommand\footnotesize{%
   \@setfontsize\footnotesize\@xipt\@xipt
   \abovedisplayskip 10\p@ \@plus2\p@ \@minus5\p@
   \abovedisplayshortskip \z@ \@plus3\p@
   \belowdisplayshortskip 6\p@ \@plus3\p@ \@minus3\p@
   \def\@listi{\leftmargin\leftmargini
               \topsep 6\p@ \@plus2\p@ \@minus2\p@
               \parsep 3\p@ \@plus2\p@ \@minus\p@
               \itemsep \parsep}%
   \belowdisplayskip \abovedisplayskip
}
\makeatother

\usepackage[american]{babel}

\usepackage[babel]{csquotes}

\usepackage[authordate, natbib, noibid, isbn=false, backend=biber]{biblatex-chicago}  
\usepackage{multicol}

\usepackage{authblk}

\usepackage[bottom]{footmisc}

\usepackage{amsmath}
\usepackage{amsfonts}

\usepackage{caption}
\usepackage{geometry} 
\usepackage{setspace}
\usepackage{float}

\usepackage{booktabs}

\usepackage{etoolbox}
\newcommand{\zerodisplayskips}{%
  \setlength{\abovedisplayskip}{0pt}%
  \setlength{\belowdisplayskip}{12pt}%
  \setlength{\abovedisplayshortskip}{0pt}%
  \setlength{\belowdisplayshortskip}{12pt}}
\appto{\normalsize}{\zerodisplayskips}
\appto{\small}{\zerodisplayskips}
\appto{\footnotesize}{\zerodisplayskips}

\usepackage{graphicx}

\usepackage{booktabs} 

\usepackage{array}

\usepackage{paralist} 

\usepackage{verbatim}

\usepackage{subfig} 

\usepackage{fancyhdr} 
\usepackage[nottoc, notlof, notlot]{tocbibind} 
\usepackage[titles, subfigure]{tocloft} 

\usepackage{comment}

\begin{document}

\thispagestyle{empty}

\bigskip

\begin{center}
	{\Large Uncertainty in Grid Data: A Theory and Comprehensive Robustness Test}
\end{center}

\bigskip


\begin{center}
{\large Akisato Suzuki}

\begin{singlespace}
School of Law and Government\\
Dublin City University\\
Dublin 9, Ireland\\
\end{singlespace}
\&\\
\begin{singlespace}
School of Politics and International Relations\\
University College Dublin\\
Belfield, Dublin 4, Ireland\\
\end{singlespace}

\begin{singlespace}
akisato.suzuki@dcu.ie\\
ORCID: 0000-0003-3691-0236
\end{singlespace}
\end{center}

\bigskip

\begin{center}
{\large Working paper\\ (February 7, 2022)}
\end{center}


\begin{singlespace}
	
\begin{center}
\textbf{Abstract}
\end{center}
	
\noindent
This article makes two novel contributions to spatial political and conflict research using grid data. First, it develops a theory of how uncertainty specific to grid data affects inference. Second, it introduces a comprehensive robustness test on sensitivity to this uncertainty, implemented in R. The uncertainty stems from (1) what is the correct size of grid cells, (2) what is the correct locations on which to draw dividing lines between these grid cells, and (3) a greater effect of measurement errors due to finer grid cells. My test aggregates grid cells into a larger size of choice as the multiple of the original grid cells. It also enables different starting points of grid cell aggregation (e.g., whether to start the aggregation from the corner of the entire map or one grid cell of the original size away from the corner) to shift the diving lines. I apply my test to \textcite{Tollefsen2012} to substantiate its use.

\end{singlespace}



\newpage

\section{Introduction}
Over the last decade or so, political and conflict research has significantly developed geo-referenced data and analytical technique for such data \autocite[e.g.,][]{Lee2019, Kikuta2022, Pickering2016, Schutte2014, Shaver2019, Sundberg2013, Tollefsen2012}. The gist of geo-referenced data is to disaggregate macro-level data into micro-level ones. For example, drought in one location might have an impact on the probability of political violence in that particular location but not on the probability of political violence in the other locations even within the same country \autocite{Theisen2011, VonUexkull2016}. Thus, when we are interested in causal effects in locality, we should resort to disaggregated, geo-referenced data rather than country-level data.

One of the universal formats in geo-referenced data is grid cells \autocite{Tollefsen2012}. In grid data, the entire map of interest (e.g., the African continent) is divided into the artificial grid cells of a certain size. The result is an artificial, apolitical division of the map. Each observation corresponds to a grid cell specified by a certain square kilometer. As grid cells are typically created in a smaller size than the entire territory of the average country, grid data allow researchers to focus on the local relationship between a causal factor and an outcome \autocites[for examples of applied research, see][]{Buhaug2011, Linke2017, OLoughlin2012, Ruggeri2017, Schutte2011, Theisen2011, Wood2015}.

However, using grid data is not without costs; we must consider uncertainty specific to grid data. There are three sources of such uncertainty.

First, there is often ambiguity in what is the correct size of grid cells. Is the correct size $3,000km^2$, $12,000km^2$, or $28,000km^2$? It is often difficult to theoretically pinpoint the only correct size (if any).

Second, there is also ambiguity in what is the correct locations on which to draw dividing lines between grid cells. In theory, it is possible to draw the first dividing line from any point of the latitude and longitude. What are the correct locations of the dividing lines to make each grid cell correctly include causally related variables and separate causally unrelated variables?

Third, all else equal, finer grid cells are more prone to measurement errors; an event recorded in a grid cell might have actually took place in a nearby grid cell. This is because it is often difficult to identify the exact zone of the phenomenon of interest, practically or conceptually \autocite{Kikuta2022}. For example, it is practically difficult to identify the exact location of political violence in a remote area that is difficult for researchers or journalists to reach. For another example, it is conceptually challenging to draw a clear line between a conflict area and a non-conflict area within a country.

As known, a measurement error affects statistical and causal inference \autocite{Pearl2010}, while the first two sources of uncertainty specific to grid data are related to the so-called Modifiable Areal Unit Problem: the choice of a geographic unit affects inference \autocite{Lee2019, Lee2020a, Openshaw1983, Soifer2019}. For political geographic units, this problem can sometimes be addressed based on a theoretical justification; for example, the norm of national sovereignty may be a good justification for why nation-states are a plausible unit for the study of foreign policy. In grid data, on the other hand, because grid cells are by definition \textit{apolitical} geographic units -- the point that is usually celebrated as an advantage of grid data \autocite[365]{Tollefsen2012}, it is difficult to find the same kind of theoretical justification \autocite[107-108]{Soifer2019}.

This article has two objectives. One is to develop a theory of exactly how each of the three sources of uncertainty specific to grid data affects inference. While the existing literature sometimes explores empirically how changing the size of grid cells affects statistical inference \autocite[e.g.,][]{Ito2020, Lee2020a, Linke2017, Schutte2011}, this article provides a theoretical foundation on why it is necessary to use different grid cell specifications for robustness checks. A specific choice of grid cell specification affects statistical and causal inference \autocite[95]{Soifer2019}. Thus, results from a single grid cell specification must not be taken for granted (unless that specification is the only theoretically conceivable way, which is unlikely in social science), without sensitivity to different specifications being examined.

The other objective is to introduce a comprehensive robustness test to accommodate the three sources of uncertainty specific to grid data \autocites[for a general principle of robustness tests, see][]{Neumayer2017}. First, it aggregates grid cells into a larger size of choice as the multiple of the original grid cells. Second, in the process of aggregation, it also allows for different starting points of grid cell aggregation, to diversify the locations on which to draw the dividing lines between aggregated grid cells. For example, if the grid cell size is to be doubled, the starting point of the aggregation can be either from the corner of the entire map or one grid cell away from the corner. This creates a shift in the dividing lines between the grid cells of the aggregated size, by one grid cell of the original size. Third, by doing the above two, measurement errors can also be accommodated, in that if the results are robust regardless of plausible aggregations, it suggests the measurement errors are not serious.

This article is also accompanied by a new R package (co-developed with Johan Elkink) to make the robustness test easy to implement. While \textcite{Pickering2016} introduces software to generate the dataset of grid cells of any size, which is useful to create a new dataset based on one's research purpose, our R package can be used for any existing grid data as long as the row and column numbers of the grid cells are available. This is useful particularly when the original source of data is unavailable and only a grid data format is available (e.g., in the case of replication analysis). The R package also enables streamlining a robustness check efficiently without the necessity of creating multiple datasets of different grid cell sizes in advance.

To exemplify the use of the robustness test, I apply it to a previous study that, using a PRIO-GRID dataset \autocite{Tollefsen2012}, concludes the null effect of drought on the likelihood of the onset of civil armed conflict \autocite{Theisen2011}. Varying the size of grid cells and the starting point of grid cell aggregation, I find greater uncertainty than the original study showed.

In the rest of the article, I first discuss further how the three sources of uncertainty in grid data affect inference. Second, I explain how aggregating grid cells can serve as a comprehensive robustness test on possible sensitivity in statistical estimation using grid data. Third, I exemplify the robustness test by showing how to vary the size of grid cells and the starting point of grid cell aggregation in the application, and present the results from the test. Finally, concluding remarks are stated.

\section{How the specification of grid cells affects inference}
In this section, I explicate the three sources of uncertainty specific to grid data, in greater detail. The first is the uncertainty over what size of grid cells we should use to divide the entire map. Practical limitations often define the smallest possible grid cell size. But there is an infinite number of ways, in theory, to aggregate the smallest grid cells because a distance is a continuous measure. Should one side of the cell be 50km, 51km, 100km, or 125km? Practically, we may think in a more discrete way, such as 50km, 100km, 150km, and so on. No matter how we aggregate gird cells, the point is that the size of grid cells directly affects statistical and causal inference.

\begin{figure}[H]
  \includegraphics[scale=1]{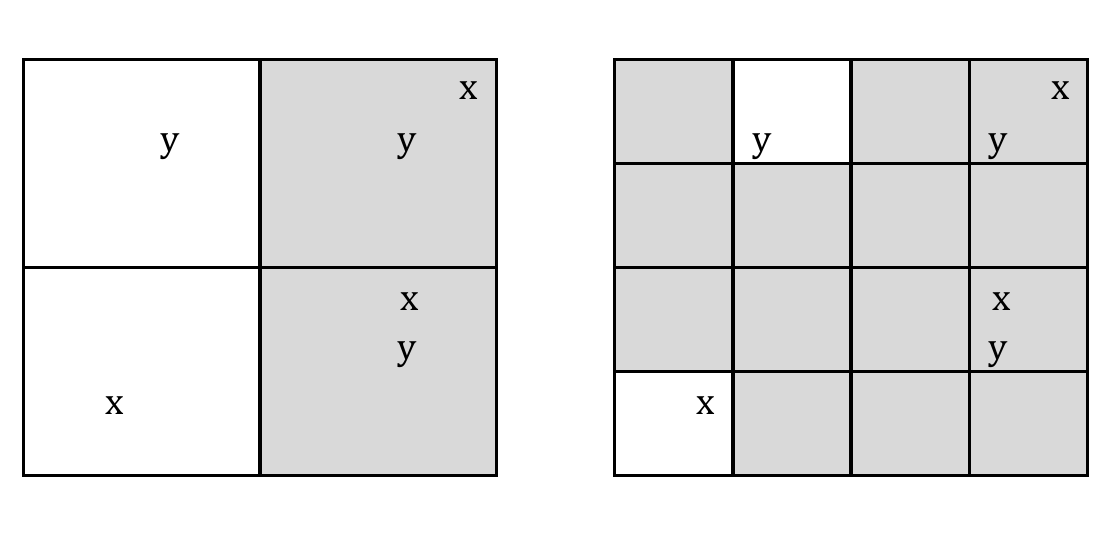}
  \centering
  \caption{How different grid cell sizes affect statistical and causal inference. x: a binary treatment variable taking a value of one; y: a binary outcome variable taking a value of one. Shaded cells: either $X=1, Y=1$ or $X=0, Y=0$.}
  \label{diffgrid}
\end{figure}

Figure \ref{diffgrid} is a crude but illustrative example -- an approach similar to the explanation by the canonical MAUP study \autocite{Openshaw1983} but fine-tuned for grid data in particular. In the left panel, the entire map is divided into four grid cells, while in the right panel, divided into sixteen grid cells. ``x'' and ``y'' represent a value of one for the binary treatment and outcome variables $X$ and $Y$ (e.g., the presence of drought and that of political violence). In the left panel, $X$ is positively correlated with $Y$ in two out of the four grid cells, while in the right panel, it is in fourteen out of the sixteen grid cells (see the gray shaded cells -- note that the empty cells are where neither the treatment nor the outcome is present, i.e., $X=0, Y=0$). In other words, the degree of the positive association between $X$ and $Y$ is stronger in the right panel than in the left panel. This is of course an artifact of grid cell specifications, because the entire map is the same in both panels. In other words, what size of grid cells is used to divide the entire map influences statistical and causal inference.

We could justify a particular size of grid cells based on a theory of how far a causal effect can go to affect an outcome. If grid cells were larger than what such a theory warrants, it would risk an ecological fallacy (i.e., creating a statistical association between the treatment and outcome in the area where the actual causal effect does not exist). If grid cells were smaller than what the theory warrants, it would risk something opposite: failing to detect a statistical association between the treatment and outcome in the area where the effect actually does exist.

\begin{figure}[H]
  \includegraphics[scale=1]{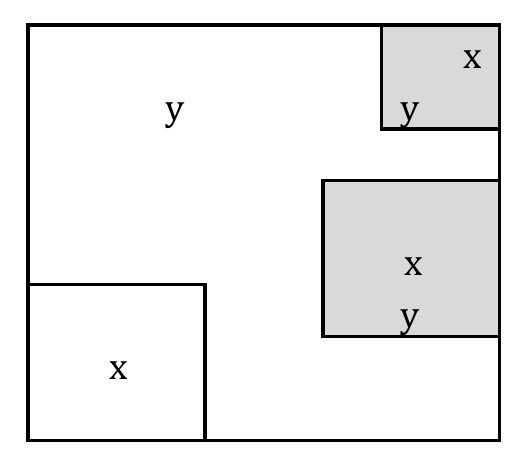}
  \centering
  \caption{A way to divide the map theoretically, which does not divide the map evenly and is not apolitical. x: a binary treatment variable taking a value of one; y: a binary outcome variable taking a value of one. Shaded cells: either $X=1, Y=1$ or $X=0, Y=0$.}
  \label{asymgrid}
\end{figure}

However, even if the size of grid cells could be justified theoretically, inference could still be affected by the second source of uncertainty specific to grid data -- the ambiguity in what locations are correct to draw the dividing lines between the grid cells. This question is impossible to theoretically answer, because grid cells are by definition supposed to be apolitical and therefore atheoretical units. In other words, any dividing lines that are drawn by grid cells must not have any political or theoretical meaning attached, unlike national borders that represent the political/theoretical meaning of sovereignty. In terms of Figure \ref{diffgrid}, even if we could theorize the treatment variable can have a causal effect only on the immediate neighboring area, we could not create grid cells based on that like in Figure \ref{asymgrid}, as the dividing lines of grid cells are defined by the theoretical causal effect of the treatment and no longer apolitical. Of course, spatial data like in Figure \ref{asymgrid} are plausible to use to examine the spatial effect of the treatment on the outcome, but are not what grid data are supposed to be. Grid data are supposed to consist of the grid cells of equal size to divide the entire map, which makes the data apolitical and atheoretical.

How is statistical and causal inference affected by a possible variation in the locations of the dividing lines? Let us keep assuming that the treatment variable can have a causal effect only on the immediate neighboring area. There are two difficulties dividing the map into the grid cell of equal size to reflect this theoretical point.

\begin{figure}[H]
  \includegraphics[scale=1]{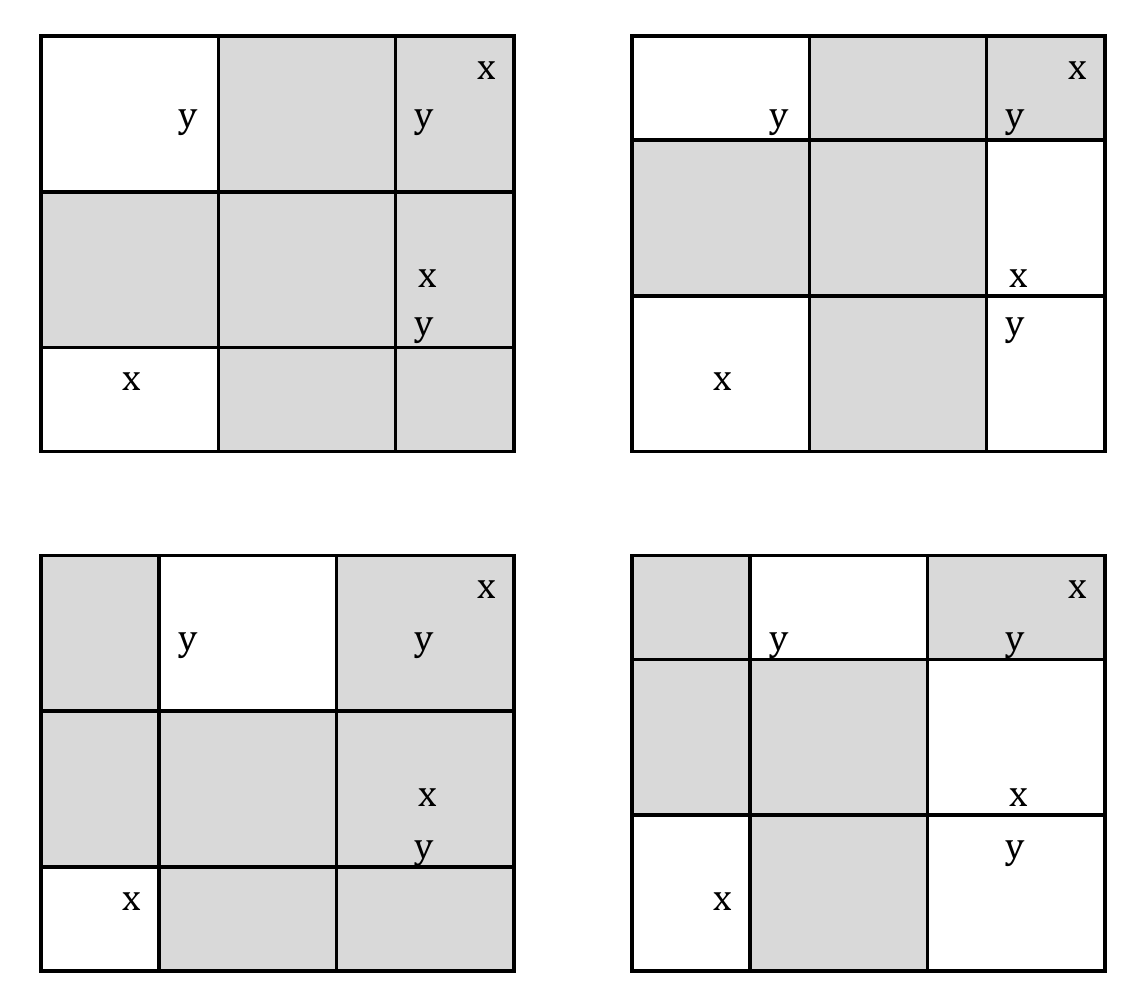}
  \centering
  \caption{How the variation in the locations of the dividing lines of grid cells affects statistical and causal inference. x: a binary treatment variable taking a value of one; y: a binary outcome variable taking a value of one. Shaded cells: either $X=1, Y=1$ or $X=0, Y=0$.}
  \label{diffline}
\end{figure}

First, not all grid cells can have the same size because the theoretically correct size of grid cells cannot divide the entire map of the current example evenly (it is indeed very unlikely that the correct grid cell size can divide the real map, which is much more complex than the square used here, evenly everywhere). Figure \ref{diffline} is four possible ways to divide the map. Since grid cells do not have (and must not have) any political or theoretical meaning attached, there is no way to decide which way is ``correct'' \textit{a priori}.

Second, depending on how to draw lines to place the correct size of grid cells, the positive correlation between $X$ and $Y$ becomes stronger or weaker. For example, theoretically there are two cases where $X$ are causally positively related to $Y$ from a \textit{theoretical} point of view (one on the top right corner and the other on the middle right corner). Only the left top and bottom panels of Figure \ref{diffline} correctly capture both causal relationships, \textit{given this specific configuration of the locations of the variables}. However, again, because grid cells are supposed to have no political or theoretical meaning attached, we cannot or should not decide what is the correct location of each grid cell, based on the location of the phenomenon of interest. In short, the apolitical and atheoretical nature of grid data makes it difficult to use only one type of grid-cell specification (by ``grid-cell specification,'' I mean a way to divide the entire map into the grid cell of equal size).

The third source of uncertainty in grid data is a measurement error. For example, the codebook of the PRIO Conflict Site \autocite[3]{Hallberg2011} states:

\begin{quote}
A drawback with circular conflict zones is that they cover more territory than is actually affected by the conflict, including territories of neighboring countries. Due to the nature of armed conflict it may be impossible to gain information on the exact locations of armed encounters, occupied territories, and rebel bases.
\end{quote}

\noindent
A measurement error over the location of an event poses a dilemma about how small the size of grid cells should be. The smaller grid cells are, the more detailed micro phenomenon we can examine on the one hand, but the greater the impact of a measurement error on the other hand.

\begin{figure}[H]
  \includegraphics[scale=1]{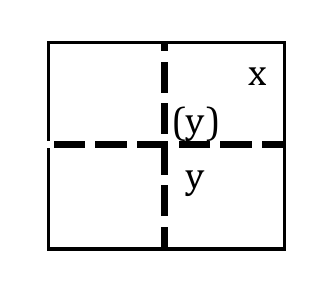}
  \centering
  \caption{Example of a tradeoff between a measurement error and a grid cell size. x: a binary treatment variable taking a value of one; y: an observed binary outcome variable taking a value of one with a measurement error; (y) is the true but unobserved location of the outcome.}
  \label{meserror}
\end{figure}

Figure \ref{meserror} clarifies why a smaller grid cell size increases the impact of a measurement error on inference. ``x'' is a binary treatment variable taking a value of one, ``(y)'' is the true location of an outcome event but \textit{unobserved}, and ``y'' is an \textit{observed, therefore measured} binary outcome variable taking a value of one. ``y'' is recorded in a location different from ``(y),'' meaning it is measured with an error. If the size of grid cells is set smaller as indicated by the dashed lines, the relationship between $X$ and $Y$ is different from the true state. If a larger grid cell size indicated by the solid line (i.e., the aggregate of the four small cells) is used, the relationship between $X$ and $Y$ is correctly captured even though ``y'' is measured with an error.

\section{A comprehensive robustness test by the aggregation of grid cells}
To summarize the previous section, there are three sources of uncertainty in grid data: the size of grid cells, the locations on which to draw the dividing lines between grid cells, and a greater effect of measurement errors due to finer grid cells. A comprehensive robustness test for grid data need to capture all three sources of uncertainty.

To do so, I propose aggregating grid cells into several different sizes and varying the starting point of aggregation. The aggregation of grid cells enables diversifying the sizes of grid cells and the locations of the dividing lines. A theory might be unable to determine the single plausible size of grid cells, but it could at least determine a plausible \textit{range} of grid cell size \autocite[104]{Soifer2019}. To reflect such a range, we can create the separate datasets of the plausible grid cell sizes.

\begin{figure}[H]
  \includegraphics[scale=0.75]{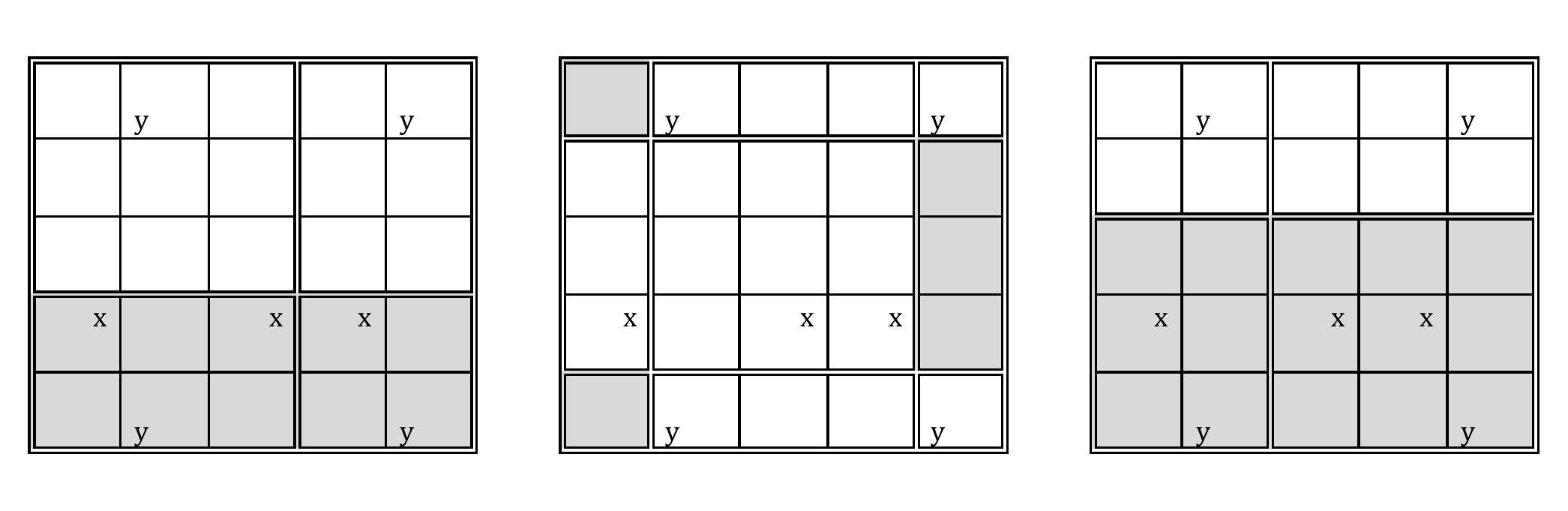}
  \centering
  \caption{Example of grid cell aggregation with the shifting of the starting point. The signle lines denote the original grid cell size. The double lines denote the aggregated grid cell size. x: a binary treatment variable taking a value of one; y: a binary outcome variable taking a value of one. Shaded cells: either $X=1, Y=1$ or $X=0, Y=0$.}
  \label{gridshift}
\end{figure}

The aggregation also allows us to change the locations on which dividing lines are drawn to place grid cells, even when we are using existing grid data that have the predefined positioning of grid cells. We can shift the starting point of aggregation. For example, if we aggregate the data by multiplying the original grid cell size by three, we can create three different versions of the data, where the shift of the aggregation starting point from the north west to the south east is zero, one, and two (see Figure \ref{gridshift}).

By doing the above two, measurement errors can also be accommodated. This is because, if the results are robust regardless of plausible aggregation, it suggests the measurement errors are not serious.

In short, if the estimates do not differ among the datasets of different and theoretically plausible grid cell specifications, it implies the estimates are robust. If they do differ, it encourages the researcher to further investigate why they differ. Is it because some of the grid cell sizes, and/or some of the ways to draw the dividing lines, are less appropriate than others? Is it because key variables are prone to measurement errors? Whether the results remain robust or not, the comprehensive robustness test provides more statistical information than a statistical model using the dataset of only one grid cell specification, to encourage further research. The following section exemplifies the use of the test.

\section{Application}
In this section, I apply my test to \textcite{Theisen2011}.\footnote{Their replication dataset can be obtained from the Peace Research Institute Oslo website at https://www.prio.org/publications/5109 (accessed on January 7, 2022).} \textcite{Theisen2011} argued against the conventional wisdom that drought increases the likelihood of the onset civil armed conflict, based on the statistical models using the grid data of the African continent, the PRIO-GRID data \autocite{Tollefsen2012}. It found the measure of drought not statistically significant and its point estimate negative rather than positive.

I aggregate the grid cells in the original data from the north west edge of the map. In the aggregation process, each variable used in the original model must also be aggregated in a theoretically consistent way \autocite[for the details of the variables used, see][93--96]{Theisen2011}. The cell-specific binary variables (the onset of conflict, drought, marginalized ethnic groups, and the capital city) are aggregated to take a value of one, if at least one of the grid cells to be aggregated observes a value of one. The cell-specific continuous variables (a grid cell's distance to a border and a grid cell's population) are aggregated to take the mean of the observed values in the grid cells to be aggregated. The country-specific variables (an infant mortality rate, the autocracy-democracy scale, and conflict history) are aggregated to take the mean of the observed values in the grid cells to be aggregated. In other words, if all those grid cells belong to one country, the mean is the same as the values of each of these cells. If the grid cells to be aggregated belong to different countries, the mean is the average of the values between/among these countries. If a variable takes a missing value for any grid cell to be aggregated, the cells containing missing values are ignored when the aggregated value is computed; if all grid cells to be aggregated to a larger one have missing values, the aggregated value is also missing.

The original grid cell size is approximately $55km \times 55km = 3,025km^2$, the one defined by the PRIO-GRID data \autocite{Tollefsen2012}. In the aggregation process, it is important to consider the theoretically plausible maximum size. In particular, when the treatment and outcome variables are binary and their maximum value (which is $1$) is used for aggregated grid cells, aggregating grid cells into an arbitrarily large size almost surely results in the strong positive correlation between them. If the maximum grid cell size is not theoretically justified, the correlation cannot be considered causal.

I multiply the side of the original grid cell of \textcite{Theisen2011} by two to six; thus, the plane of the aggregated grid cell becomes the square of the side multiplied by one of these values. In other words, the smallest aggregated cell size is the default $(55km \times 2)^{2} = 12,100km^2$ while the largest is $(55km \times 6)^{2} = 108,900km^2$. The longest straight line within the square is the diagonal, $a\sqrt2$, where $a$ is the length of one side; the diagonal of $108,900km^2$ is approximately $467km$.

I reason that the grid cell size of $108,900km^2$ is theoretically plausible, as follows. If a car moves 20km per hour on average, it takes 23.35 hours to go all the way through the diagonal of $467km$. In reality, a car may not be able to go straight through the diagonal and at the constant rate of 20km per hour. But, even if we assume it takes three times longer, it is still 70.05 hours to cover from one corner to the other corner within the grid cell of $108,900km^2$. In the data, the temporal unit of analysis is years. People could move over that distance during well below the time period of one year, if they were affected by drought and decided to move for water and survival. Such migration could increase the likelihood of conflict \autocite{Reuveny2007}. Rebel groups could also move over as long a distance, if they were interested in recruiting people who were affected by drought and had a lower cost of participating in armed rebellion \autocite[86]{Theisen2011}. Finally, even though a few countries are smaller than $108,900km^2$ in terms of their land mass (e.g., Rwanda or Burundi), rebel groups often operate across borders as well.

When the grid cells are aggregated, I also vary the starting point of aggregation. I shift the grid cell from which the aggregation begins, by from one grid cell of the original size up to the aggregated grid cell size minus one. For example, if I multiply a grid cell size by three, I create three different versions of the data, where the shift of the aggregation starting point from the north west to the south east is the zero grid cell of the original size (i.e. no shifting), the one grid cell of the original size, and the two grid cells of the original size.

\textcite{Theisen2011} used a random subsample of 5\% of the observations where there is no onset of civil armed conflict, most probably following the recommendation by \textcite{King2001a}. \textcite{King2001a} point out that when the outcome variable is a rare event, it does not affect the estimation much even when most of no-event observations are dropped, and in the current case, it also helps reduce spatial autocorrelations. Thus, I follow this practice as well (the aggregation of grid cells is done before taking a subsample). I generate 30 random subsamples, making sure the results are not driven by a particular subsample; the reason for choosing 30 subsamples is simply to show a possible variation in the estimates but also keep the plots of the results legible. To estimate the effect of drought on the likelihood of civil armed conflict onsets, I use logistic regression as done in the original study.

\begin{figure}[H]
  \includegraphics[scale=.175]{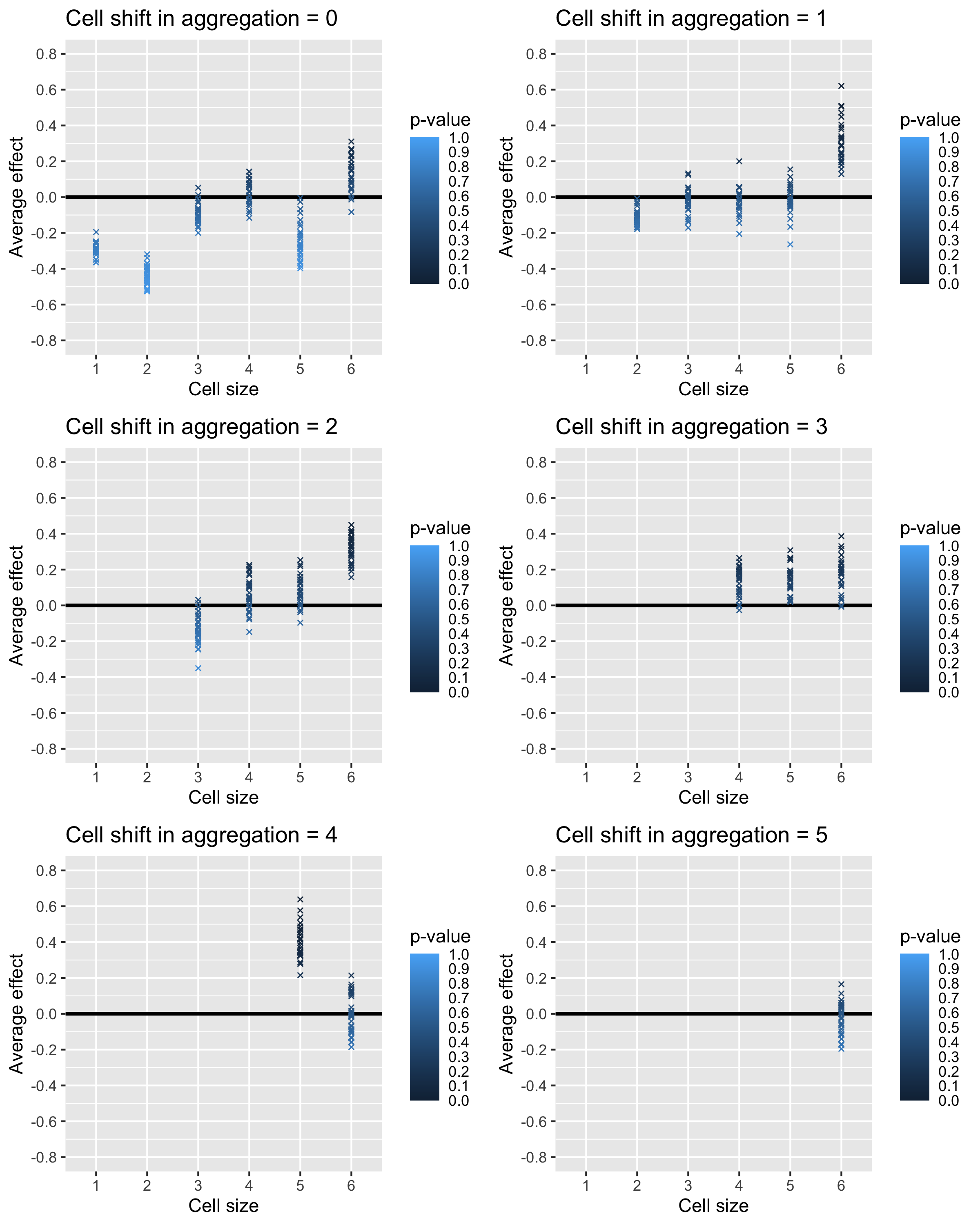}
  \centering
  \caption{Point estimates of the average effect of drought across 30 subsamples. Darker color of the symbol means a smaller p-value.}
  \label{resultTHB1}
\end{figure}

Figure \ref{resultTHB1} presents the results. The y-axis is the point estimate of the average effect of drought on the likelihood of the onset of civil armed conflict on the log odds ratio scale.\footnote{Different grid cell sizes create the different numbers of observations and change the baseline likelihood of the onset of civil armed conflict while keeping such an onset rare events (less than $0.5\%$ of the observations) even for the largest grid cell size used here. Therefore, the comparison of effect sizes across models are more meaningful on the log odds ratio scale, the relative scale of an effect, than on the probability scale, the absolute scale of an effect.} If the color of the symbol (\texttimes) is darker, it means a smaller one-tailed p-value for the point estimate being a positive value greater than zero. The x-axis indicates grid cell sizes. As there are 30 subsamples per grid cell specification, there are 30 point estimates.

There is a large variation in terms of both points estimates and p-values, across different grid cell sizes and shifts. When I use the original grid cell size (cell size $= 1$) or a small aggregated size such as a cell size of two or three, almost all point estimates of the average effect of drought are negative and the p-values tend to be large. Thus, in these grid cell specifications, the finding of \textcite{Theisen2011} holds.

The difference appears if the grid cells are aggregated to larger sizes. The point estimates tend to indicate greater effects of drought on the likelihood of the onset of civil armed conflict, and the p-values tend to become smaller. This result empirically points to uncertainty in the estimate of the causal effect over the different grid cell sizes. Interestingly, this tendency does not apply to the cell size of six when the shift in aggregation is either four or five. This point empirically illustrates uncertainty in the estimate of the causal effect over the different locations on which the dividing lines between grid cells are drawn.

\begin{figure}[H]
  \includegraphics[scale=.175]{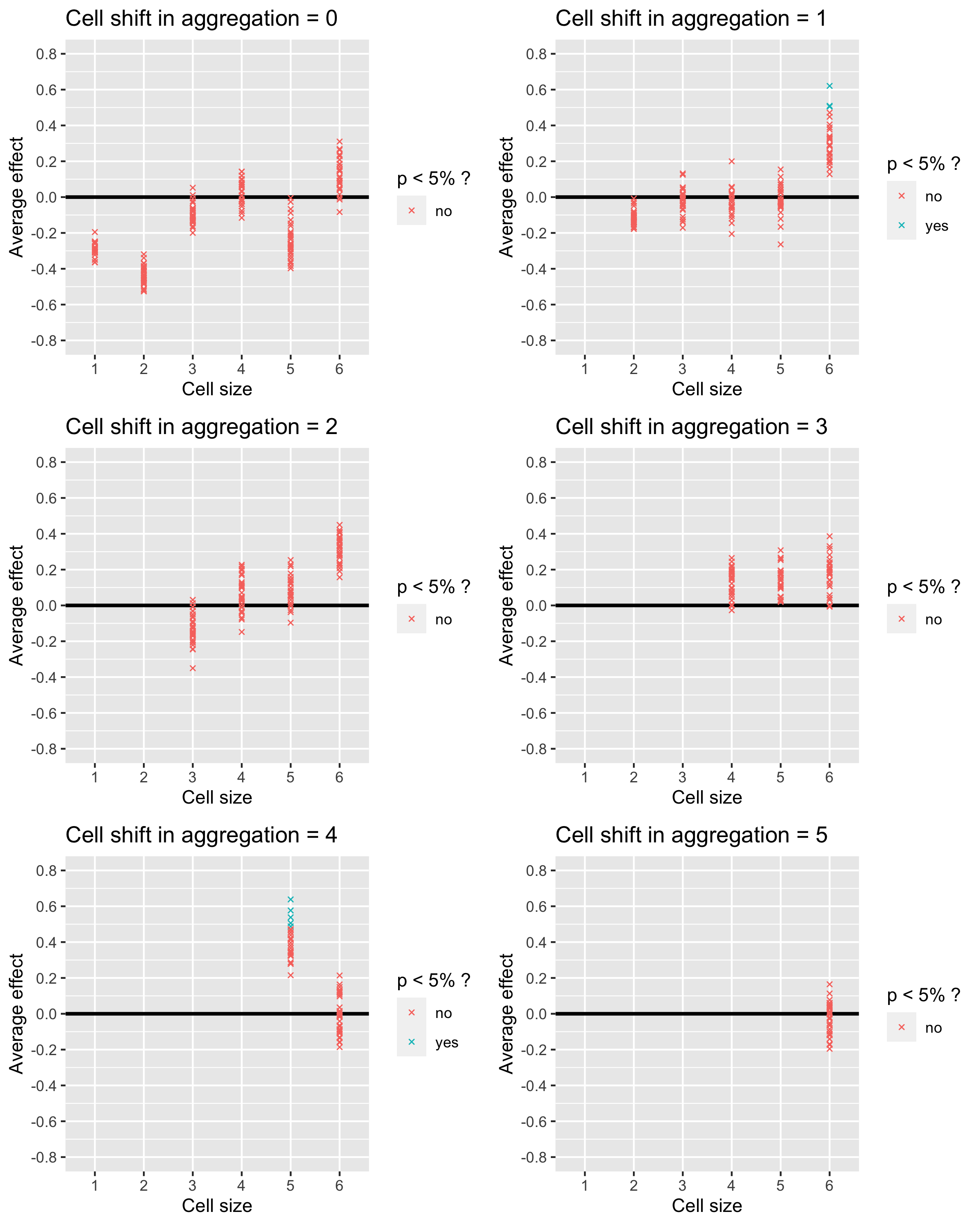}
  \centering
  \caption{Re-coloring the symbols by statistical significance/non-significance in Figure \ref{resultTHB1}.}
  \label{resultTHB2}
\end{figure}

For the Null Hypothesis Significance Testing, I re-color the symbols based on statistical significance vs. non-significance at the threshold of $p<5\%$ (see Figure \ref{resultTHB2}). There are many more models that produce statistical non-significance than those that produce statistical significance. Thus, from this point of view, my test suggests the original conclusion is fairly robust. Yet, it is controversial whether the Null Hypothesis Significance Testing is a plausible way to evaluate the uncertainty of an effect \autocite[see][]{Amrhein2019, Gelman2011, Suzuki2022}. If we take a p-value as a continuous measure of uncertainty \autocite{Lew2012}, the results here suggest a large variation in uncertainty over the average effect of drought on the likelihood of the onset of civil armed conflict. From this point of view, the original results of \textcite{Theisen2011} seem sensitive to the choice of grid cell sizes and the starting point to aggregate the grid cells. Whether researchers take either viewpoint of statistical inference, my test is useful to examine sensitivity in statistical estimation using grid data.

\section{Conclusion}
I have shown how different aggregations of grid cells (i.e., their sizes and the starting point of aggregation) bring the variation and, therefore, uncertainty, in statistical estimation. I suggest researchers should not rely only on the results from only one grid cell specification, unless they can theoretically justify the use of that particular grid cell specification. If there is a significant variation in the estimates across the datasets of different grid cell specifications, such as the one this article has found, researchers should consider why different grid cell sizes produce different results, and the results from which grid cell size are theoretically more plausible than the others. If they cannot identify any theoretical reason to put greater credibility on specific grid cell specifications while facing a significant variation in the estimates, they can only conclude the results are inconclusive \autocite{Gross2015, Kruschke2018a}.

Greater theorization on the geography of politics aligns well with the recent trend in a more focus on connecting a theory to a statistical model \autocite{Pearl2018a, Keele2020, Lundberg2021}. It is also possible that greater theorization will lead researchers to conclude grid cells are not an appropriate unit of analysis. For example, in the case of conflict research, the artificial borders of grid cells might not reflect theoretical borders relevant to conflict dynamics, such as the terrains that facilitate or hinder the movement of armed groups \autocite{Fearon2003}. In such a case, political or theoretical geographic units of analysis might be preferred.


\section*{Acknowledgments}
I would like to thank seminar participants at DCU for their helpful comments, and Johan Elkink for encouraging me to independently develop his initial idea on this topic. The accompanying R package,``rbstgrid,'' has been co-developed with Johan Elkink at the School of Politics and International Relations, University College Dublin, and will become available at \url{https://akisatosuzuki.github.io/programs.html}. I would like to acknowledge the receipt of funding from the Irish Research Council (the grant number: GOIPD/2018/328) for the development of this work. The views expressed are my own unless otherwise stated, and do not necessarily represent those of the institutes/organizations to which I am/have been related.


\section*{Supplemental online material}
The R code to replicate the analysis will become available at \url{https://akisatosuzuki.github.io/papers.html}.

\printbibliography

\end{document}